\documentclass[runningheads]{llncs}

\usepackage{graphicx}

\newcommand{\cond}[1]{\emph{(#1)}}

\begin{document}

\title{Diversity by Design: Balancing Protection\\ and Inclusion in Social Networks\thanks{All authors contributed equally to this paper.}}

\titlerunning{Diversity by Design: Balancing Protection and Inclusion in Social Networks}

\author{Paula Helm\inst{1} \and
Loizos Michael\inst{2,3} \and
Laura Schelenz\inst{1}}

\institute{
International Center for Ethics in the Sciences and Humanities\\
University of Tübingen, Tübingen, Germany\\
\email{\{paula.helm, laura.schelenz\}@uni-tuebingen.de}
\and
Open University of Cyprus, Nicosia Cyprus\\
\and
CYENS Center of Excellence, Nicosia, Cyprus\\
\email{loizos@ouc.ac.cy}
}

\maketitle


\begin{abstract}
The unreflected promotion of diversity as a value in social interactions --- including technology-mediated ones --- risks emphasizing the benefits of inclusion at the cost of not recognizing the potential harm from failing to protect stigmatized or marginalized individuals. Adopting the stance that technology is not value-neutral, we attempt to answer the question of how technology-mediated social platforms could accommodate \emph{diversity by design}, by balancing the often competing values of protection and inclusion. This short paper presents our research agenda as well as initial analysis and outcomes. Building on approaches from scenario planning and the methodology of Value Sensitive Design, we identify ethical principles and arguments on how to curate diversity, which we seek to operationalize through formal argumentation.
\end{abstract}


\section{Diversity in Social Media, Too Much of a Good Thing?}

While modern digital technologies hold an enormous potential to bring together people with different skills, experiences, opinions, backgrounds, competencies, and resources, helping transcend geographic boundaries in connecting people, the unchecked adoption of diversity can end up becoming a hindrance to meaningful online engagement. Unsurprisingly, then, designing systems that strike a fine balance between the protection and the inclusion of individuals is a challenging task, with existing social media and networking platforms drawing criticism both for \cond{i} failing to take advantage of the diversity of their users, and \cond{ii} failing to curtail the negative sides of diversity, such as abusive behavior \cite{Schelenz.2021}. 

In order to tackle this criticism from an interdisciplinary Ethics and Computer Science perspective, we explore the socio-technical possibilities of intervening in a diversity-aware chatbot to empower users. Our guiding research question is: \emph{Can we build algorithms to support meaningful user interactions on social networking platforms in a way that considers the benefits and risks of diversity?}

We attempt to answer this question in the context of the EU-funded project ``WeNet -- The Internet of Us'' (\texttt{https://www.internetofus.eu}), where the goal is to facilitate machine-mediated diversity-aware social interactions. As part of the project's ongoing pilot studies, a chatbot application has been developed that supports users --- university students, in the current pilots --- across geographic, social, and cultural backgrounds in seeking help from the ``crowd'' of other platform users through written messages, with the expectation that the platform will be considerate of sensitivities and power asymmetries. 

The design of the platform and the chatbot follows the ``Values in Design'' research program \cite{FriedmanHendry.2019,KnobelBowker.2011}, which opposes the widespread attitude of technology being value-neutral. Accordingly, the project seeks to carefully identify the values to be inscribed in the developed technologies, and what implications this process will have when materialized in the real world. In this spirit, this work considers the use-case of a chatbot to reflect on and discuss our posed question.

\section{Ethical Diversity Curation when ``Asking for Help''}

In society, diversity is often invoked to create associations with pluralism, tolerance, or simply ``good'' policy. Diversity has become popular as a marketing strategy and been adopted in the rhetoric of big tech companies \cite{Chi.2021}. Yet, diversity is more complicated than usually insinuated in public discourses. From an ethical perspective, it is important to critically review diversity concepts and pay attention to the negative implications of unfiltered diversity. 
While diversity in technology-mediated spaces does not necessarily differ from diversity in society, the mediation raises questions about the responsibility of technology designers and providers to curate diversity. This is similar to political actors' responsibility to, e.g., protect minorities in liberal democracies \cite{Waldron.2012}. Even though popular technology may be designed by a number of powerful private actors, the quality of exchange in platforms today resembles the one in public spaces. It is therefore ethically relevant to regulate online and communication spaces to prevent harm to users and allow for democratic opportunities to negotiate these spaces \cite{Crawford.2016}. 

As per the WeNet project, we understand diversity as a descriptive and normative concept. In the first direction, diversity is a concept that helps us define differences between users and ultimately classify elements of their being (e.g., skills and practices) that can complement other users' characteristics to form a moment of mutual support. 
In the second direction, we treat diversity as a value and differentiate between ideas of diversity as intrinsic value (diversity is important for itself) and diversity as instrumental value (diversity helps achieve other values such as inclusion or productivity) \cite{ZimmermanBradley.2019}. This differentiation is relevant for the ethical intervention in the diversity-aware WeNet chatbot and platform. There, we build on our understanding of diversity as an instrumental value that supports inclusion, tolerance, and justice (cf.\ Section~\ref{method}). However, diversity also needs to be weighted against other demands such as protection and efficiency. 

Furthermore, diversity can be seen as a strategy, which nicely aligns with our understanding of diversity as an instrumental value. In recommender systems, for example, diversity has been established as the practice of diversifying the item sets recommended to users in order to increase user satisfaction \cite{Miyamoto.2018}, or foster democratic principles \cite{Vrijenhoek.2021}. Helberger discusses diversity in the context of media and news recommendations, and argues that diversity is a goal but also a design strategy to enable different choices and hence increase autonomy of users \cite{Helberger.2011,Helberger.2019}. In that context, the term ``Diversity by Design'' is used to refer to ``the idea that it is possible to create an architecture or service that helps people to make diverse choices'' \cite{Helberger.2011}. We adopt the term ``Diversity by Design'' in the title of this paper and expand further the idea to include not only diverse choices, but also safe, autonomous, and beneficial interaction in our envisioned chatbot. 

\begin{figure}[t]
{}\hfill\includegraphics[width=0.9\textwidth]{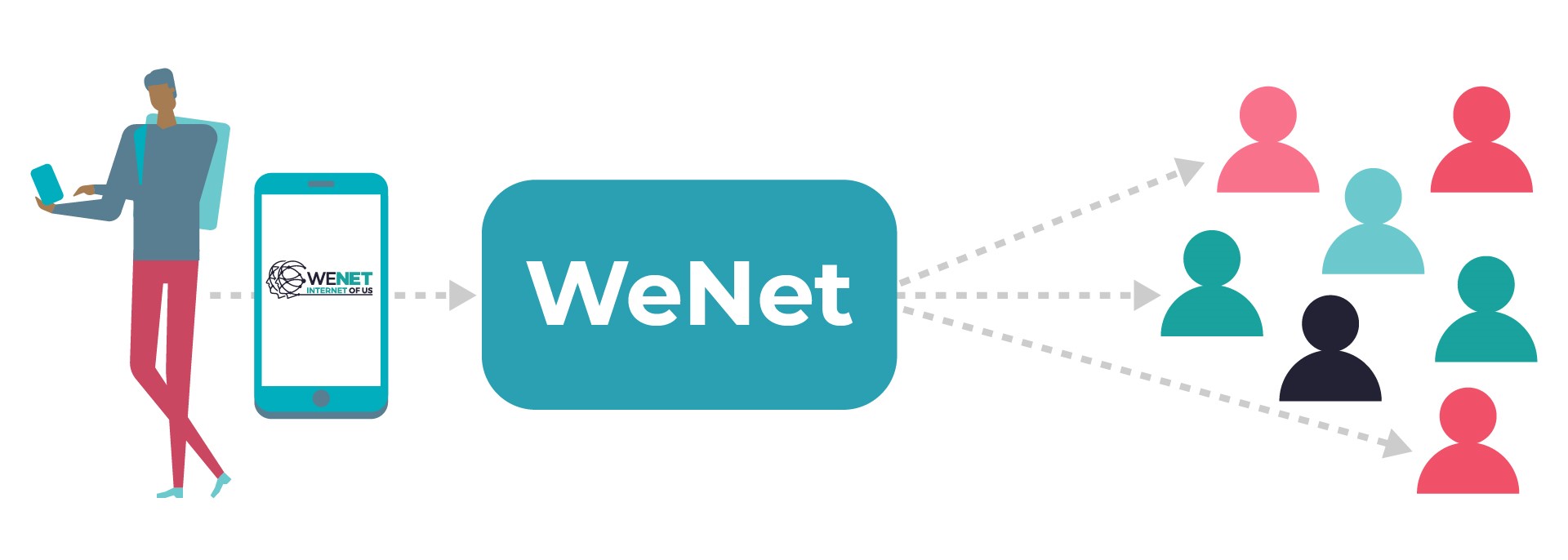}\hfill{}
\caption{Visual representation of the WeNet technology: A user can ask any question to a diverse community of peers, through a machine-powered social platform.}
\label{figure1}
\end{figure}

The chatbot ``Ask for Help'' is designed as an instant messaging application within Telegram, running over the WeNet diversity-aware platform (cf.\ Figure~\ref{figure1}). The chatbot allows users to send questions (or ``requests'') to the community of registered users of the chatbot. When a user poses a question to the community via the command ``/question'', they are requested to specify whether they want to ask users who are ``similar'' to them or who are ``different''. While this request initially served research purposes to define a better understanding of users' needs for diversity, it also allows users to specify the scope of respondents. The queries by users are analyzed algorithmically on the basis of previously collected user profiles in an attempt to determine the most appropriate users to respond.

In this step, the system relies on the descriptive conceptualization of diversity (see the discussion above); data has been collected through self-report mechanisms as well as mobile sensor and geolocation tracking to determine a user's set of skills and practices. This sociological concept of diversity goes beyond classic demographic approaches to building user profiles and thus may enrich the user's experience with the chatbot's matching to other users; cf.\ \cite{Schelenz.2021}. Ideally, the chatbot thus takes into account aspects of both fit (diversity of skills and practices; i.e., who has the skills to provide an answer to my request?) and individual preferences for diversity (i.e., do I need users with similar experiences to provide for a safe space, or can I benefit from expertise that complements my own?).

The ethical intervention in the chatbot was developed after a first round of experiments revealed that students were asking sensitive questions about mental health, on the one hand, but also covering highly controversial topics around the COVID-19 pandemic or religion, on the other hand. To mitigate potential harm to users from, e.g., hate speech and abuse, or insensitive responses that provoke a deterioration of vulnerable users' mental health, the ethics and design team decided to introduce extra algorithmic measures and design instruments, such as nudges and report mechanisms, to offer users more opportunities for protection. 

\section{Methodology: Value-Sensitive Scenario Analysis}
\label{method}

In our research, we combine approaches from the field of scenario planning \cite{Dean.2019,Schoemaker.1995}, with the methodology of Value Sensitive Design (VSD) \cite{FriedmanHendry.2019}. Through these techniques and methods, we seek, iteratively, and in an agile manner, to:

\begin{itemize}

    \item[\cond{1}] conceptualize the relevant and desirable values, and identify the potential tensions between them, for ethically accommodating diversity in the chatbot;
    
    \item[\cond{2}] empirically determine, primarily through scenario thinking, how the relevant values are weighted against each other in case of incompatibilities;
    
    \item[\cond{3}] technically implement the outcomes of the earlier phases by operationalizing the identified values and their weights in a formal representation language.

\end{itemize}

As part of phase~\cond{1}, we approach diversity as a normative value specifying ``what ought to be'', and also as an instrumental value to be leveraged to achieve other important values that empower users and benefit the society: promoting numerous opinions and political orientations, achieving equity and social justice \cite{Crenshaw.1988}, expanding experiences and knowledge, fostering tolerance and mutual understanding \cite{UNESCO.2005}, and advancing business goals such as solving complex tasks by bringing together diverse team members in productive ways \cite{Tugend.6November2018}.

Juxtaposed with these benefits of diversity are several risks: perceiving social signals as incoherent noise, harming or silencing minorities or marginalized groups through hate speech \cite{Maitri.2009}, and being non-conducive to meaningful connectivity by excluding safe spaces for sensitive issues such as health  \cite{Helm.2018}. Intriguingly, sacrificing a certain level of diversity (e.g., in expressed opinions) to avoid the risks above, ends up promoting and protecting diversity (e.g., by supporting minorities to express freely), which is key to equality in liberal societies \cite{Waldron.2012}. 

\begin{figure}[t]
{}\hfill\includegraphics[width=0.9\textwidth]{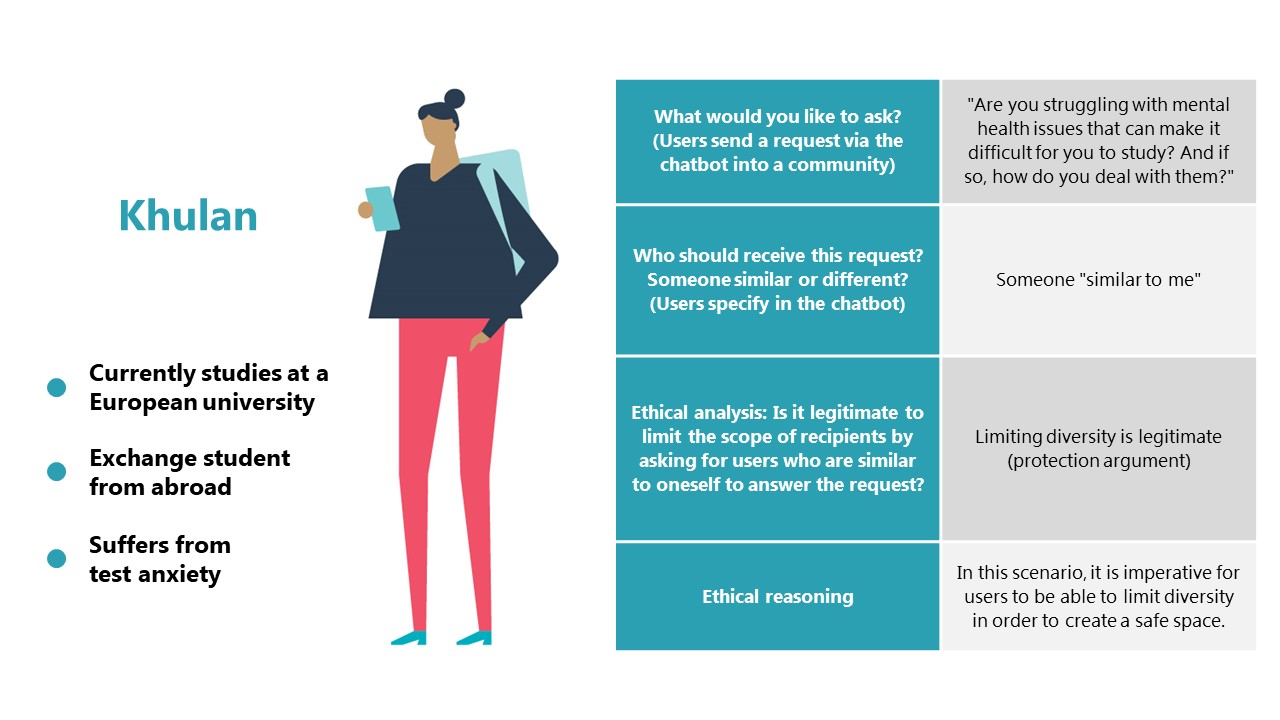}\hfill{}
\caption{An example scenario  which stands for requests made through the ``Ask for Help'' chatbot that are sensitive and may require a safe space. In this particular scenario, we determine that limiting diversity, i.e., the scope of possible respondents to the user's question, is legitimate. Of course, this means that the system discriminates against some other chatbot users by excluding them from being considered as potential respondents, but the need for protection of the user asking the question in this particular case outweighs the risks of exclusion of others from consideration as respondents.}
\label{figure2}
\end{figure}

As part of phase~\cond{2}, we empirically determine and carefully assess the contexts in which diversity unfolds, and whether its limitation should be promoted, avoided, or dismissed as unethical. Inspired by real user profiles and requests from the WeNet project's first pilot, we draw on methods of scenario analysis \cite{York.2020} to develop four scenarios and supplement them with two fictional ones to explore the fringe cases of potential misuse. Each scenario (cf. Figure~\ref{figure2}) includes a user's background and request, and whether they chose to limit the respondents.

Furthermore, drawing on the common distinction between instrumental and fundamental values \cite{ZimmermanBradley.2019}, with the latter being intrinsically valuable and, therefore, weighing more heavily than the former, we identified various values touched upon in the scenarios. The values are: \cond{instrumental values} inclusion, tolerance, freedom of choice, efficiency; \cond{fundamental values} autonomy, well-being, health, dignity, justice. Loosely inspired by Walzer's analysis in ``Spheres of Justice'' \cite{Walzer.1984}, we also take into account the contextual dependency of value judgments, especially when it comes to balancing conflicting values at tension with each other. 

Ideal-typically, we distinguish three spheres to which the different scenarios relate. In analyzing the relative relevance of the sometimes competing values, we take the different requirements (protection and inclusion) corresponding to the different spheres into account. We are aware that the boundaries of the spheres we present in this framework are fluid in reality and in fact often difficult to clearly delineate, but for our analysis they provide the necessary analytical basis to make a reasoned decision in ambivalent cases. The spheres are: \cond{maximum freedom sphere} leisure / sports / art; \cond{shared resources sphere} economy, politics, education; \cond{protection / sensitive sphere} religion, health / medicine, psychology.

Finally, in addition to the various values and the social spheres, we consider the economic-historical-cultural situatedness from which users make inquiries and from which these inquiries are answered \cite{Haraway.1988}. The six scenarios are analyzed along these three considered dimensions (i.e., values, social spheres, and situatedness) to determine how diversity should be curated in each case and why.

\section{Operationalizing Ethics through Formal Argumentation}

When the first iteration of phases~\cond{1}--\cond{2} of our value-sensitive scenario analysis is completed, we can proceed, as part of phase~\cond{3}, to operationalize our findings. 

We approach this operationalization by utilizing formal argumentation \cite{Dung.1995}, turning the identified conditions under which diversity should be curated into arguments, and the relative weights that result from the considered values into an attacking relation between the arguments. Applying standard formal argumentation semantics helps decide how competing values that might be promoted in some context are to be handled when determining how to curate diversity.

The preliminary arguments stemming from our initial ethical analysis are: 

\begin{description}
    
    \item[\textbf{Efficiency Argument:}] It is useful and it is not unethical to limit diversity if the request speaks to a certain demographic, or involves knowledge / skills tied to a specific social practice.
    
    \item[\textbf{Protection Argument:}] It is imperative to limit diversity if the request relates to the protection / sensitive sphere or is sensitive due to a cultural / social context.
    
    \item[\textbf{Inclusion Argument:}] It is unethical to limit diversity if the request relates to the shared resources sphere.
    
    \item[\textbf{Freedom-of-Choice Argument:}] It is possibly problematic but it is not strictly unethical to limit diversity if the request relates to the maximum freedom sphere even if it might be exclusive.
    
    \item[\textbf{No-Harm Argument:}] The question of curating diversity is irrelevant if the request is violating a fundamental ethical principle at any rate.

\end{description}

Once a determination is made on which argument prevails and how diversity is to be curated, various instruments can be used to implement the associated curation action: nudging users to revise their request, ensuring that users have a sufficiently-rich set of options on limiting the respondents to their requests, allowing users to block received requests deemed intrusive or offensive, or even offering a mechanism to launch complaints or report unethical requests. 

Our ongoing work as part of phase~\cond{3} involves the use of the Web-STAR system \cite{Rodosthenous.2019} as a substrate to formally encode the aforementioned scenarios and arguments. Our working hypothesis is that the use of an argumentation-based representation and reasoning system will help us make concrete the relevant aspects of contexts and how those relate to the premises of the arguments, and to support, if needed, a further iteration of the value-sensitive scenario analysis.

Ultimately, we envision this line of work to produce a knowledge-based system that automatically determines how to ethically curate diversity in social networking contexts such as those of our chatbot use-case. The system will receive information about the user making a request and the request itself, expressed in some structured form or a controlled natural language \cite{Kuhn.2014}, and will provide transparency on the decisions made. Through Machine Coaching \cite{Michael.2017,Michael.2019} the system will facilitate its in-situ ethical elaboration and fine-tuning \cite{Michael.2020}, while accommodating the encoding of suggestions on which of the instruments could be more appropriately used to implement each diversity curation action.

\section{Limits of Ethical Intervention and Looking Ahead}

While our research agenda centers on the need to ethically align diversity in social networking technology, we recognize the limits of ethical intervention, especially when this intervention is expressed and encoded algorithmically. Contemporary Ethics goes, in fact, to great lengths to emphasize the importance of context for moral judgments and to admonish us not to fall into a kind of simplistic binary thinking that pigeonholes the (social) world into an ``either/or'' logic, thus failing to recognize its indeed messy, intersectional, and fluid nature \cite{Law.2007}. 

While our framework of analysis attempts to accommodate these convictions as much as possible by taking into account different social spheres and the economic-historical-cultural situatedness of users, we are still bound by the demands of automation. Machine learning and reasoning tools used to develop the chatbot require us to \cond{i} establish rules that are generalizable and formalizable, and \cond{ii} enable outcomes that are clear enough to be translated into code. 

So, is the ``Ethics by Design'' project not, thus, a contradiction in terms? It is definitely not possible to pursue this project without compromises, and probably not without mistakes \cite{Bijker.2017}. However, since values are inscribed in technology anyway and since neutrality is an illusion \cite{Friedman.1996}, our approach makes the inscription of values with all its constraints as considerate, conscious, and transparent as possible. The use of formal argumentation further supports the extension of the inscribed values in an elaboration tolerant manner, if and when the need arises.


Looking ahead, this paper has set a research agenda for the technical implementation of ethical arguments and considerations to promote diversity in social media and networking platforms. The next steps include the refinement of scenarios and ethical arguments as well as the formal codification of the arguments. Additional next steps will also need to address the specific technical, aesthetic, social, and policy instruments that can be used to curate diversity in terms of both balancing and promoting inclusion and protection in social networks.

As we move forward, we expect that the work will likely require considering additional scenarios due to the number of value tensions arising from a combination of values. 
While in this paper we have limited ourselves to a selection of scenarios that are related to the issue of weighing protection against inclusion, other important value tensions include, for example, freedom and justice, or privacy and transparency. We will further focus on other issues derived from empirical data from the anticipated second pilot of the WeNet project.


\paragraph{Acknowledgements:}

This work was supported by funding from the EU's Horizon 2020 Research and Innovation Programme under grant agreements no.\ 739578 and no.\ 823783, and from the Government of the Republic of Cyprus through the Deputy Ministry of Research, Innovation, and Digital Policy. The authors would like to thank Martel Innovate for the graphic art, and the anonymous reviewers of the KI 2021 Workshop on AI and Ethics for their valuable feedback. 


\bibliographystyle{plain}
\bibliography{references}

\end{document}